\documentstyle[prb,aps]{revtex}
\begin{document}
\title{Apparent Metallic Behavior at $B=0$ of a Two Dimensional Electron System
in AlAs}
\draft

\author{S. J. Papadakis and M. Shayegan}
\address{Department of Electrical Engineering, Princeton University, Princeton,
NJ, 08544}
\date{Feb 10, 1998}
\maketitle
\begin{abstract}
We report the observation of metallic-like behavior at low temperatures
and zero magnetic field in two dimensional (2D) electrons in an AlAs quantum
well.  At high densities the resistance of the sample decreases with
decreasing temperature, but as the density is reduced the behavior changes to
insulating, with the resistance increasing as the temperature is decreased.
The effect is similar to that observed in 2D electrons in Si-MOSFETs, and in 2D
holes in SiGe and GaAs, and points to the generality of this phenomenon.
\end{abstract}
\pacs{PACS:  71.30.+h}

The question of whether or not a metal-insulator transition can
occur in two dimensions at zero magnetic field has been of great recent
interest.  Using scaling arguments, Abrahams et al.\cite{Abrahamsscaling}
showed that a non-interacting 2D carrier system with any amount of disorder
will be localized at zero temperature.  Subsequent experiments provided
evidence to support this theoretical prediction.\cite{BishopTsui} However,
recently, various investigators\cite
{KravMI,Henini94,Ural95,PopovicMI97,ColeridgeMI97,Lam97,Hanein97,MYSimmons97} 
have discovered 2D carrier systems that show metallic behavior.  They
observe that for a range of 2D densities ($n$), the resistivity of their
samples decreases by nearly an
order of magnitude as the temperature ($T$) is decreased.  When $n$ is reduced
below a critical density ($n_c$) they observe a transition to insulating
behavior.  One
difference between the earlier and the more recent experiments is that the new
samples have a much higher carrier mobility ($\mu$).  One theoretical
investigation asserts that this higher $\mu$ combines with a broken inversion
symmetry in the confining potential well to allow spin orbit effects to create
the metallic state,\cite{PudalovMI97} while another
hypothesizes that the higher $\mu$ allows for stronger electron-electron
interaction, and that this interaction causes the metallic
state.\cite{DobrosavljevicFinkelshtein}  However, there is still no clear
model supported by experimental results to explain this metallic behavior.

So far, the metallic behavior has been observed in 2D electron systems (2DESs)
in Si-MOSFETs,\cite{KravMI,PopovicMI97} 2D hole systems (2DHSs) in
GaAs/AlGaAs heterostructures\cite{Henini94,Ural95,Hanein97,MYSimmons97} and
SiGe quantum wells (QWs),\cite{ColeridgeMI97,Lam97}  and now in a 2DES in
AlAs.  To provide an overview, and for further discussion, some of the
important parameters of these systems are shown in Table \ref{table1}.  AlAs is
an interesting material for 2DESs because it combines some of
the properties of GaAs with those of Si.  Since it is grown in
the same molecular beam epitaxy (MBE) systems as GaAs samples, very clean
samples can be fabricated.  However, it is similar to Si in that the minima of
the AlAs conduction band are at the X-points of the Brillouin Zone.  (The
minima in Si are near the X-points.)  In addition, in AlAs QWs like ours, which
are grown on the (100) surface of the GaAs substrate, one can cause the
electrons to occupy the conduction band ellipsoids either perpendicular to or
parallel to the plane of the 2DES by varying the width of the
QW.\cite{vandeStadt95,vanKesteren89}  Our data indicate that in our sample,
only {\em one} of the {\em in-plane} ellipsoids is occupied.
By patterning Hall bars along and perpendicular to the direction of the
occupied ellipsoid's major axis, we are able to measure the resistance along
these two directions.  We observe anisotropy in the measured
resistance, and the data along both directions show metallic behavior at high
$n$ and insulating behavior at low $n$. 

Our sample was grown by MBE on an undoped GaAs (100) substrate. The 2DES is
confined to a 150 \AA-wide AlAs QW which is separated from the dopant atoms
(Si) by a 300 \AA-wide front barrier of Al$_{0.45}$Ga$_{0.55}$As and a rear
barrier consisting of 25 periods
of a GaAs/AlAs (10.5 \AA/8.5 \AA) superlattice.  On a sample from
this wafer, we patterned two Hall bars oriented perpendicular to each other
(L-shaped) along the [001] and [010] directions.  The Hall bars were patterned
by standard photolithographic techniques and a wet
etch.  Ohmic contacts were made by alloying AuGeNi in an N$_{2}$ and H$_{2}$
atmosphere for 10 minutes.  A front gate of 350 \AA\ Au on top of 50 \AA\ Ti
was deposited on top of the active regions of the Hall bars to control $n$.

Our $T$ dependence measurements were performed in a pumped $^{3}$He
refrigerator at $T$ from 0.28 K to 1.4 K.  We measured $T$ using a
calibrated RuO$_{2}$ resistor.  We used the standard low-frequency AC lock-in
technique with an excitation current of 1 nA to measure the four point
resistance of the sample.  The data were taken by fixing the front-gate voltage
($V_g$), and measuring both the longitudinal ($R_{xx}$) and
transverse ($R_{xy}$) resistances as a function of perpendicular magnetic field
($B$).  These magnetoresistance measurements were used to determine $n$ at that
$V_g$.  Gate leakage was monitored throughout the experiment and it never
exceeded 10 pA.  The $T$ was then raised to 1.4 K and continuously lowered back
to the base $T$ (0.28 K) over a period of three hours.  The densities and $T$
dependencies were repeatable at the same gate voltages.
The results of $R_{xx}$ and $R_{xy}$ magnetoresistance measurements at 0.28 K
for $V_g = 0$ are shown in Fig. \ref{1}.  Note the high
quality of the data, with the appearance of Shubnikov-de Haas 
oscillations at a field as low as 0.6 T and the fractional quantum Hall effect
at Landau level filling factor $\nu = 2/3$ as well as at $\nu = 1/3$ (see Ref.
\onlinecite{Lay93}).  

Before presenting the $T$-dependence data, we will describe some of the
characteristics of our AlAs 2DES. Several observations lead us to believe that
in our sample only {\em one in-plane} conduction-band ellipsoid is occupied.
First, previous cyclotron resonance (CR) measurements on samples from this
wafer reveal a CR effective mass $m_{CR} = 0.46m_{e}$.\cite{Lay93}  This mass
is in excellent agreement with the expected CR mass if in-plane ellipsoids are
occupied.\cite{Adachinote}  It is very different from $m_{CR} = m_t = 0.19m_e$
which would be observed if an ellipsoid perpendicular to the plane were
occupied.  Second, the data of Fig. \ref{1} show minima in $R_{xx}$ and
plateaus in $R_{xy}$ for both even and
odd filling factors, and the Shubnikov-de Haas oscillations show no beating.
Moreover, the two $R_{xx}$ traces from the two perpendicular Hall bars show an
anisotropy in $\mu$.  We conclude from these observations that only one of the
two in-plane ellipsoids is occupied:  the magnetoresistance data suggest that
there is only one occupied subband, while the $\mu$ anisotropy indicates that
the Fermi surface in the plane of the 2DES can be anisotropic, consistent with a
single in-plane ellipsoid being occupied.  It is possible that a slight angular
deviation from the ideal growth direction could account for the lifting of the
expected degeneracy of the in-plane ellipsoids, as a splitting of only a few meV
would be sufficient to produce a single occupied
subband.\cite{Lay93,TPSmithnote}  

We now compare the characteristics of the 2DES in our sample with
those of Si-MOSFETs.  First, the conduction band ellipsoids in bulk Si and AlAs
are comparable, with similar values for $m_l$ and
$m_t$.\cite{AndoFowlerSternnote}  However, in
contrast to our sample, the Si-MOSFET 2DESs which have been studied so far
occupy {\em out-of-plane} ellipsoids.  As a result, transport in the plane is
isotropic with an effective mass $m_t = 0.19m_e$.  The mobilities at base
$T$ of our sample for the trace shown in Fig. \ref{1} ($n = 2.08 \times
10^{11}$ cm$^{-2}$) are 6.1 m$^{2}$/Vs for the high-$\mu$ direction and 4.2
m$^{2}$/Vs for the low-$\mu$ direction.  The highest mobilities we measure,
for the highest density ($n = 2.73 \times 10^{11}$ cm$^{-2}$), are 7.7
m$^{2}$/Vs and 4.7 m$^{2}$/Vs.  These
mobilities are comparable to the highest mobilities reported for Si-MOSFET
2DESs (see Table \ref{table1}).\cite{KravMI}  Finally, as seen in Fig. \ref{1}
and Ref. \onlinecite{Lay93}, our sample exhibits clear fractional quantum Hall
effect, an effect rarely seen in Si-MOSFETs.

Figure \ref{2} summarizes the $T$ dependence of the zero-$B$ resistivity
($\rho$) for a range of $n$ from $2.73 \times 10^{11}$ cm$^{-2}$  to $0.59
\times 10^{11}$ cm$^{-2}$.  The results for both high and low mobility
directions are shown.  As with other experiments,
\cite{KravMI,PopovicMI97,ColeridgeMI97,Lam97,Hanein97,MYSimmons97} the
data can be split into three regimes.  In the lowest $n$ traces, the behavior
is insulating throughout the $T$ range measured, with $\rho$ rising
monotonically as $T$ is reduced.  The highest $n$ traces show
metallic behavior throughout the $T$ range, with  $\rho$ 
decreasing monotonically as $T$ is reduced.  For intermediate $n$,
$\rho$ exhibits a nonmonotonic dependence on $T$:  it initally rises as $T$ is
lowered, shows a maximum, and then decreases with decreasing $T$.  These data
are very similar qualitatively to the results of previous experiments.  

As already mentioned, a theoretical explanation for data like these, with
experimental evidence to support it, does not exist.  We expect, though, that
the Fermi energy ($E_F$) might be an important parameter.  In our sample, $E_F$
at the highest density is 16 K and at the lowest density is 3.6 K.  Our
$T$-range in these measurements (0.28 K to 1.4 K) is only about an order of
magnitude smaller than $E_F$.  Most of the 2D carrier systems investigated so
far also have $E_F$ comparable to the $T$-range over which the experiment is
done. This raises the possibility of finite-$T$ effects causing the
metallic-like behavior.  In fact, Henini et al.\cite{Henini94} have fitted to
their GaAs 2DHS data an expression [$\mu/\mu_o \sim 1-(T/E_F)^2$] describing the
temperature dependence of screening\cite{DasSarma86}
and found that the fit was very good.  Our data is not fit well by
this equation, but we cannot rule out this effect because the exact expression
for temperature-dependent screening depends on details of disorder and material
parameters.

Another possible model has been put forth by Pudalov.\cite{PudalovMI97}  He
suggests that the metallic-like data may be fitted by an empirical dependence
\begin{equation}
\rho(T) = \rho_{o} + \rho_{1}exp(-T_{o}/T).
\end{equation}
The second term is intended to account for an energy gap caused
by a spin-orbit interaction.    For $n$ where our 2DES exhibits a metallic
behavior throughout the measured $T$-range (traces {\em a} to {\em f} of Fig.
\ref{2}), this equation fits our data well through the whole $T$-range.  The
fits are not shown in Fig. \ref{2} because they are indistinguishable from the
data.  To show the accuracy of the fits, we present representative Arrhenius
plots of ($\rho - \rho_o$) vs. $1/T$ in Fig. \ref{3}a.  For clarity, only the
curves for the high-$\mu$ direction data are shown; the curves for the
low-$\mu$ direction are very similar.  Clear exponential behavior is observed
for more than a decade of ($\rho - \rho_o$) for the highest density traces, but
as the density is reduced, the range over which exponential dependence is
observed reduces to about a factor of 5.  In Fig. \ref{3}b we show, as a
function of $n$, the values of $\rho_o$, $\rho_1$, and $T_o$ deduced from
fitting Eq. 1 to the data.  As expected, $\rho_o$ rises monotonically as $n$ is
reduced.  Also, $\rho_1$ is seen to rise smoothly and monotonically, which is
more evidence that the fits are meaningful.  The $T_o$ vs. $n$ dependence seen
in the lower part of Fig. \ref{3}b is qualitatively the same as what Hanein et
al.\cite{Hanein97} observe for their 2DHS data.  Both show a dependence that is
close to linear, and that extrapolates to $T_o = 0$ at $n = 0$.  We note that a
decreasing $T_o$ with $n$ is consistent with $T_o$ being related to spin-orbit
interaction:  the spin-splitting energy in 2D carrier systems due to interface
inversion asymmetry indeed typically decreases with decreasing 2D
density.\cite{PudalovMI97,Bychkov84}  It is also interesting to compare the
dimensionless ratio $T_o/E_F$ in our measurements to those of Hanein et
al.\cite{Hanein97}  Since both $E_F$ and $T_o$ vary approximately linearly with
$n$ in the range where the behavior is metallic, this ratio is a constant for
each experiment.  For the data of Hanein et al., $T_o/E_F \simeq 0.2$, while for
ours, $T_o/E_F \simeq 0.1.$  Despite a factor of two difference, these ratios
are similar enough to suggest that $T_o$ and $E_F$ may be important physical
parameters in all of the systems that show metallic behavior.

Taken together, the results of recent experiments make it very difficult
to overlook the anomalous low-$T$ behavior in these systems.  The similarity of
the data and the parameters from various systems, and the inability of any
current theory to describe them all, strongly suggest that there is new and
interesting physics here.  A look at Table \ref{table1} shows the similarities
among some relevant parameters.  The large values of the dimensionless
parameter $r_s$ (the interparticle spacing measured
in units of the effective Bohr radius) and of $\mu$ support the idea that
electron-electron interaction plays a role in stabilizing the metallic state.
The combination of rather small densities and large effective masses that leads
to large $r_s$ values, on the other hand, also means small values of $E_F$.
Ironically, precisely because of these small $E_F$ values, it is still
questionable if it is meaningful to infer the existence of a zero-$T$
metallic state from the available finite-$T$ data:  phenomena such as
temperature dependent screening\cite{Henini94,DasSarma86} can indeed lead to a
decrease in $\rho$ with decreasing $T$ at temperatures which are not negligible
compared to $E_F$.

In conclusion, we present data from a new and different 2DES that shows the
same metallic-like behavior and apparent metal-insulator transition recently
observed in other 2D carrier systems.  The generality of this phenomenon begs
theoretical explanation.

We would like to thank Y. Hanein, D. Shahar, D. C. Tsui, and J. Yoon
for useful discussions.  This work was funded by the National Science
Foundation.

\begin{figure}
\caption{$R_{xx}$ and $R_{xy}$ data for an AlAs 2DES ($n = 2.08 \times 10^{11}$
cm$^{-2}$) confined to the (100) plane.  $R_{xx}$ is shown along two
perpendicular ([010] and [001]) directions.   Some of the Landau level
filling factors at which the quantum Hall effect is observed are marked by
vertical lines.  The inset is a closer view of the low field data,
demonstrating the resistance anisotropy at $B$ = 0.}
\label{1}
\end{figure}

\begin{figure}
\caption{Resistivity vs. Temperature data from our AlAs 2DES.  The high (low)
mobility direction data are shown by solid (dashed) curves.
The densities, in units of $10^{11}$ cm$^{-2}$, are:  $a$:  2.73, $b$:  2.08,
$c$:  1.42, $d$:  1.22, $e$:  1.02, $f$:  0.82, $g$:  0.74, $h$:  0.72, $i$:
0.70, $j$:  0.65, $k$:  0.64, $l$:  0.63, $m$:  0.60, $n$:  0.59.  The
low-mobility curve for case $n$ is not shown because the ohmic contacts failed
at such a low density.}
\label{2}
\end{figure}

\begin{figure}
\caption{a:  Arrhenius plots of ($\rho-\rho_o$) vs. $1/T$ for traces {\em a}
through {\em f}.  Only the high-$\mu$ direction curves are shown.  b:  Values
of $\rho_o$, $\rho_1$, and $T_o$ from fits of Eqn. 1 to the data (traces {\em
a} through {\em f} of Fig. 2).  The open (closed) symbols are from
fits to the low (high)-$\mu$ direction data.  The dashed line in the lower plot
is a least-squares fit to the open circle points.  The least-squares fit to the
closed circle points (not shown) also extrapolates to very close to $T_o = 0$
at $n = 0$.}
\label{3}
\end{figure}

\widetext
\begin{table}
\caption{A summary of some of the important parameters in experiments that have
reported evidence for a metal-insulator transition in a low temperature 2DES or
2DHS at zero magnetic field.  The listed $\mu$ are the peak reported $\mu$.
Note that the different systems have a wide range of $n_c$, the density below
which an insulating state ensues, but the resistivity values at the transition
($\rho_c$) are comparable.  Also,
all of these systems have a fairly high $m^*$ and a relatively small density,
leading to small Fermi energies ($E_F$) and large $r_s$ values.  Listed here
are the ranges of $E_F$ and $r_s$ over which the experiments show a metallic
behavior.}
\label{table1}
\begin{tabular}{llrrcrrr}
&&$\mu$ (m$^2$/Vs)&$n_c$ ($10^{11}$ cm$^{-2}$)&$\rho_c$
(k$\Omega$ /sq.)\tablenotemark[1]&$m^*/m_e$&$E_F$ (K)&$r_s$\\
\tableline
2DES&Si-MOSFET\cite{KravMI,PopovicMI97}&7.1-1.0&0.85-1.7&70&0.19&4.3-15&
9.6-5.7\\
&AlAs&7.7&0.7&50&0.46&3.6-16&18-9\\
\tableline
2DHS&GaAs\cite{Hanein97,MYSimmons97}&15&0.1-0.12&~40&0.38&0.5-3.7&24-9\\
&SiGe\cite{ColeridgeMI97}&1.9&1.7&20&0.2-0.32&6-28&$\sim$
5-9\tablenotemark[2]\\
\end{tabular}
\tablenotetext[1]{The values for $\rho_c$ were determined following the method
of Ref. \onlinecite{KravMI}:  by drawing a separatrix between the metallic and
insulating curves and then extending that separatrix to $T = 0$.}
\tablenotetext[2]{This $r_s$ is a rough estimate because the value of
$\epsilon$ used for SiGe is an estimate.}
\end{table}

\end{document}